\documentclass[]{spie}  
\usepackage[]{graphicx}

\title{Solving the Imaging Problem with Coherently Integrated Multiwavelength Data} 

\author{H. R. Schmitt\supit{a,b},  T. A. Pauls\supit{a}, J. T. Armstrong\supit{a}, D. Mozurkewich\supit{c}, A. M. Jorgensen\supit{d}, R. B. Hindsley\supit{a}, C. Tycner\supit{e}, R. T. Zavala\supit{f}, J. A. Benson\supit{f} and D. J. Hutter\supit{f}
\skiplinehalf
\supit{a}Naval Research Laboratory, Code 7215, 3555 Overlook Avenue SW, Washington, DC\,20375, USA; \\
\supit{b}Interferometrics, Inc.,13454 Sunrise Valley Drive, Suite 240, Herndon, VA\,20171, USA;\\
\supit{c}Seabrook Engineering, 9310 Dubary Road, Seabrook, MD\,20706, USA;\\
\supit{d}New Mexico Institute of Mining and Technology, 801 Leroy Place, Socorro, NM\,87801, USA;\\
\supit{e}Department of Physics, Central Michigan University, Mt. Pleasant, MI\,48859, USA;\\
\supit{f}US Naval Observatory, Flagstaff Station, 10391 West Naval Observatory Road,
Flagstaff, AZ\,86001, USA
}

\authorinfo{Further author information: (Send correspondence to H. R. Schmitt)\\
E-mail: hschmitt@ccs.nrl.navy.mil, Telephone: 1 202 767 2977}

\begin{document} 
  \maketitle 

\begin{abstract}
Recovering images from optical interferometric observations is one of the
major challenges in the field. Unlike the case of observations at radio
wavelengths, in the optical the atmospheric turbulence changes the phases
on a very short time scale, which results in corrupted phase measurements.
In order to overcome these limitations, several groups developed image
reconstruction techniques based only on squared visibility and closure
phase information, which are unaffected by atmospheric turbulence. We
present the results of two techniques used by our group, which employed 
coherently integrated data from the Navy Prototype Optical Interferometer.
Based on these techniques we were able to recover complex visibilities
for several sources and image them using standard radio imaging software.
We describe these techniques, the corrections applied to the data,
present the images of a few sources, and discuss the implications of
these results.
\end{abstract}



\section{INTRODUCTION}
\label{sec:intro}  

Image reconstruction is one of the major challenges for optical
interferometry. Unlike radio interferometry, where the atmosphere
usully changes on long time scales, in optical interferometry the
atmosphere fluctuates on times scales of a few ms, resulting in
the corruption of the fringe visibility phases. Recently a lot of effort
has been devoted by several groups in the development
of imaging algorithms that use only squared visibilities ($V^2$)
and  closure phases, which are uncorrupted quantities
(see Ref.~\citenum{Buscher94,Ireland06,Lawson04} for a few examples).
However, these techniques have to combine multiple measurements,
like in the case of closure phases, so they contain less information,
resulting in higher noise.

Therefore, one of the challenges to optical interferometric imaging
is to recover as much of the phase information as possible.
Here we describe two techniques developed by our group using coherently
integrated data from the Navy Prototype Optical Interferometer (NPOI).
The first technique uses differential phases to recover the phase
information in the H$\alpha$ channel of Be stars, while the second one
applies the phase self calibration method.
These techniques allowed us to recover complex visibilities and image
sources using radio interferometric reconstruction techniques.

\section{OBSERVATIONS}
\label{sec:obs}

Our observations were taken with the NPOI~\cite{Armstrong98},
on several nights between January 2004 and May 2005. All the
observations were made with two spectrographs, simultaneously recording fringes
in 16 spectral channels in the wavelength range 560--860~nm. We observed between
one and three baselines in each spectrograph, with maximum baseline lengths
ranging from 19 to 64~m. Each star was observed several times during the night
(at least 3 scans), and their observations were interleaved with observations
of suitable calibrator stars. The NPOI records fringe frames at a rate of 2~ms,
and a typical scan lasts 30~s.

The data reductions followed 2 separate steps. First we used the standard
incoherent integration method, where the data is processed to produce $V^2$'s
averaged into 1~s points. These data are flagged to delete points with bad
pointing and fringe delay tracking~\cite{Hummel03b}, and averaged to one
spectrum per scan, which is bias corrected and calibrated. These $V^2$'s
are used to estimate the amplitudes of the complex visibilities.

The second step of the data reductions is the coherent integration of the
observations. We used the technique presented by Ref.~\citenum{Hummel03a} (see
Ref.~\citenum{Jorgensen07} and Jorgensen's contribution to this conference
for more details on improvements and other applications of this technique).
We align the 2~ms phasors and average the complex visibilities in 200~ms
subscans, to increase the S/N. These subscans are flagged and the visibility
phases are calculated.

\section{DIFFERENTIAL PHASES}
\label{sec:dif}

The differential phases are used in one of the techniques developed by our
group to correct for the effects of the instrument and the atmosphere on complex
visibilities (see Ref.\citenum{Schmitt08,Pauls06} for a detailed description of the
technique, and Ref.~\citenum{Quirrenbach99,Monnier03} for a discussion on the
differential phases method). This technique uses a priori information about
the structure of the source at one wavelength to correct the observed phases
and extrapolate this correction to the wavelength of interest. We apply this
technique to the H$\alpha$ channel of Be stars. The photosphere of these stars
is usually unresolved by our continuum observations (diameter $<$1~mas),
while the H$\alpha$ emission comes from a disk with a size of a few mas.
Since the photosphere of the star is expected to be circularly symmetric,
the intrinsic phase
of the continuum channels should be zero, making them ideal reference points for the
application of this technique. Differential H$\alpha$ phases was used by
Ref.~\citenum{Vakili97,Vakili98} to study P Cygni and $\zeta$ Tauri and
infer asymmetric structures in the emission line regions of these sources.

Of key importance to this technique is the availability of simultaneous
multiwavelength observations, making the NPOI a well suited instrument
for this application. We have that the observed phase consists of three
terms, the intrinsic phase of the
source ($\phi_0$), an atmospheric ($\phi_{atm}$) and an instrumental 
($\phi_{inst}$) components:
\begin{equation}
\label{eqn:phi}
\phi_{obs} = \phi_0 + \phi_{atm} + \phi_{inst} ~.
\end{equation}

\noindent
All terms have implicit wavelength dependencies. The idea behind this technique
is to find ways to correct the $\phi_{atm}$ and $\phi_{inst}$ contributions
to  $\phi_{obs}$, which will leave us only with the phase due to the source
($\phi_0$). In the case of the NPOI, $\phi_{inst}$ is stable, and can be
determined using calibration stars, which are circularly symmetric and should have
$\phi_0=0$. Since the effect of the atmosphere is random and additive,
one can average all the scans of calibrators observed throughout the night, and
the atmospheric term ($\phi_{atm}$) should vanish, leaving us only with the
instrumental term.

In the case of the Be stars we subtract the instrumental phase from the 
observed ones and are left with $\phi_0 + \phi_{atm}$. Since the phase in the
continuum channels of these stars is expected to be zero, or at least very
close to zero, we can determine $\phi_{atm}$ by fitting a second order polynomial
to the continuum channel phases, and interpolate over the H$\alpha$ channel.
A final correction that has to be applied to the data is the subtraction of the
amount of continuum emission contribution to the H$\alpha$ channel. Once these
corrections are done, one can calculate the complex visibilities using these
phases and the $V^2$'s from the incoherent integration,
and create a FITS file that can be used with standard radio
interferometry software (e.g. AIPS~\cite{vanmoorsel96}).

In Figure~1 we show the differential phase results of $\beta$ Lyrae~\cite{Schmitt08}.
This star is an eclipsing binary with a separation of $\sim1$~mas, where
one of the components filled its Roche lobe and is transferring material to the
other component. The gainer is surrounded by a disk of gas, which shows
strong H$\alpha$ emission~\cite{Harmanec02}.
Our observations did not fully resolve the binary system in the continuum,
however, we can see in Figure~1 that we detect the displacement of the
H$\alpha$ emission relative to the continuum photocenter. Based on these
images we can determine that the position angle of the orbit is
$\sim67.7^{\circ}$, in agreement with the position angle of the
radio emission~\cite{Umana00} and spectropolarimetry results~\cite{Hoffman98}.

  \begin{figure}
   \begin{center}
   \begin{tabular}{c}
   \includegraphics[height=17cm,angle=90]{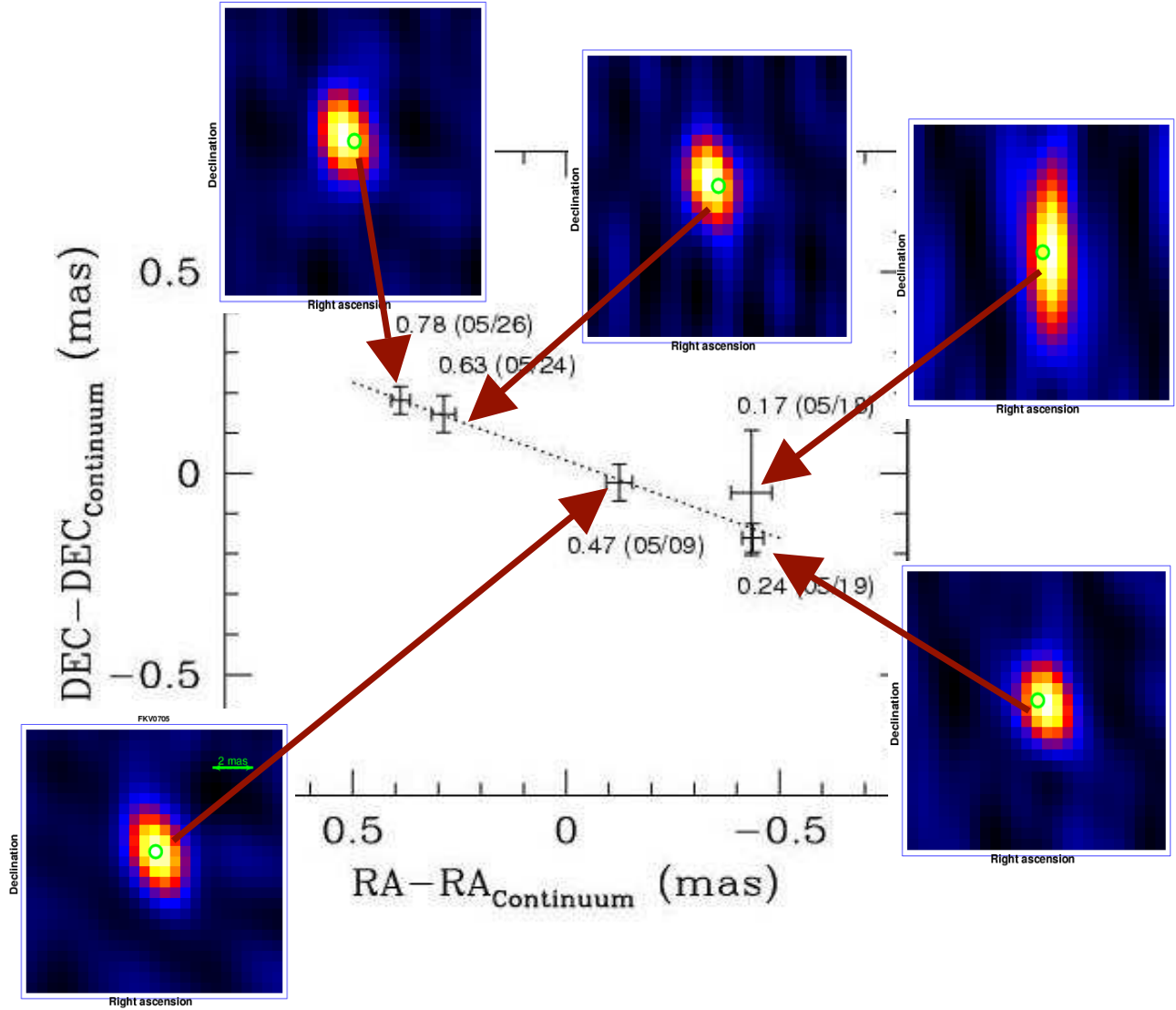}
   \end{tabular}
   \end{center}
   \caption[example]
   {Mosaic of the measured orbit of $\beta$ Lyrae, calculated by
measuring the displacement of the H$\alpha$ photocenter relative to the
continuum photocenter, and the corresponding H$\alpha$ image for each
epoch. We indicate besides each point the orbital phase and date of the
observations in 2005. The white star in the H$\alpha$ images shows the
position of the continuum photocenter. Throughout this paper we show the
deconvolving beam in the bottom left corner of each image panel. The
lowest contour level corresponds to 3$\sigma$ above the background
level, with the levels increasing by 3$\sigma \times 2^n$ steps.}
   \end{figure}

   \begin{figure}
   \begin{center}
   \begin{tabular}{c}
   \includegraphics[height=12cm]{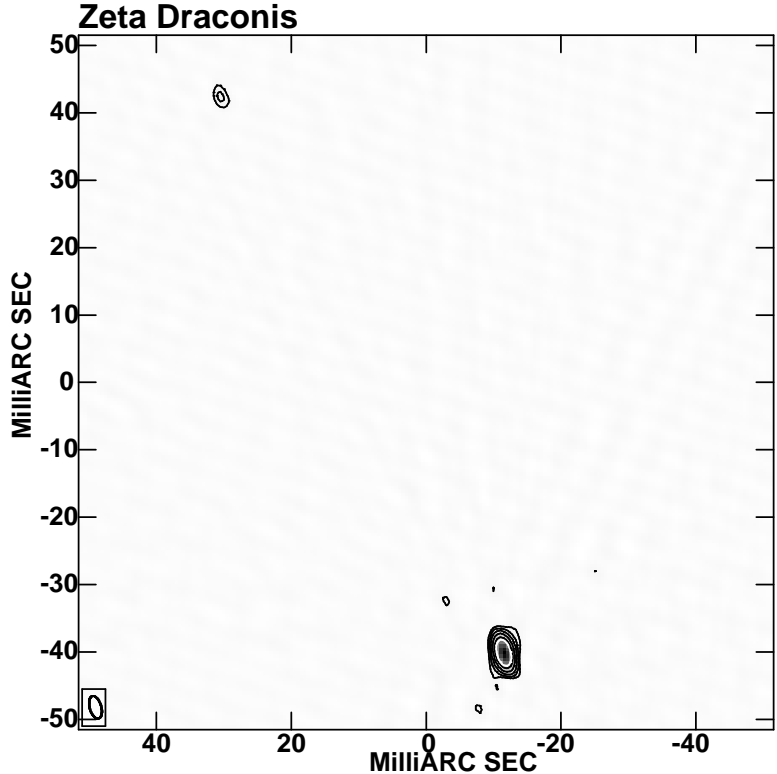}
   \end{tabular}
   \end{center}
   \caption[example]
   {First phase self-calibrated image of the binary star $\zeta$ Draconis.
Based on this image we measure $\rho=92.62$mas and $\theta=26.9^{\circ}$,
which are confirmed
with a traditional fit to the V$^2$ and triple phases of this source.}
   \end{figure}

\section{SELF CALIBRATION}
\label{sec:self}

The second technique used by our group to recover complex visibility
information from NPOI observations combines most of the processing done
for the differential phases, and self calibration~\cite{Cornwell81}.
The data reductions follow the steps described above. We use the
observed phases obtained from the coherent intergration and correct them
for the instrumental effects. This leaves us with baseline phases, which
are composed of a comnponent intrinsic to the source ($\phi_0$)
and a component due to the atmosphere ($\phi_{atm}$). These phases
are combined with the $V^2$'s, obtained from the incoherently averaged
data, to calculate the complex visibilities of the source.
These visibilities are converted to a FITS file and imported into AIPS,
where we correct the atmospheric effect on the phases using the phase
self calibration technique, widely used in radio interferometry.
A requirement for this technique to work is the simultaneous observation
of three or more baselines from which closure phases can be obtained.
This information is needed in order to remove the atmospheric effects and
recover the intrinsic source phases.

   \begin{figure}
   \begin{center}
   \begin{tabular}{c}
   \includegraphics[height=12cm]{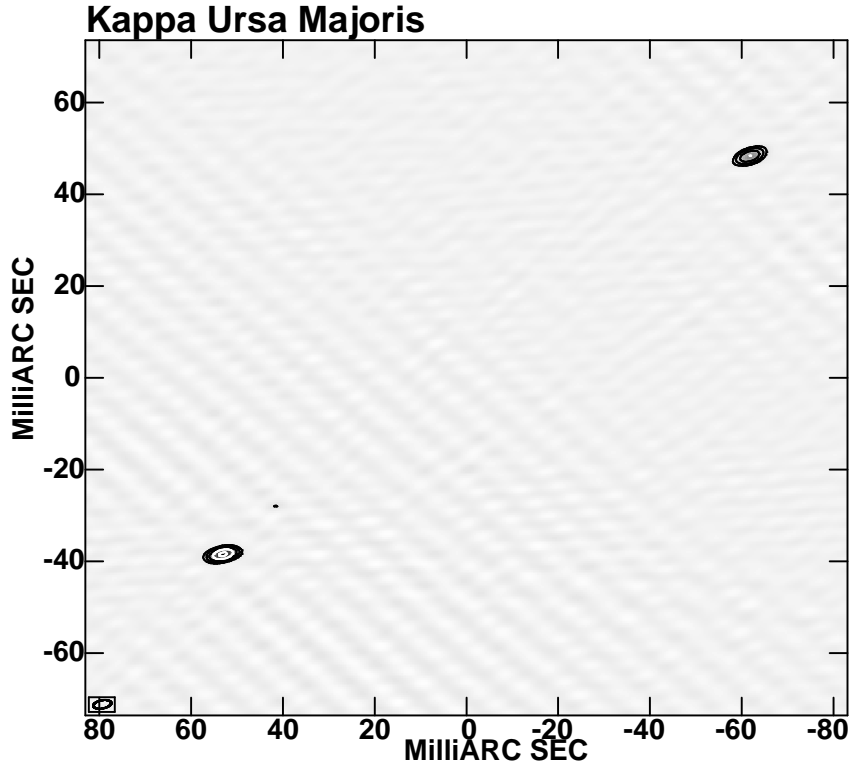}
   \end{tabular}
   \end{center}
   \caption[example]
   {Image of $\kappa$ Ursa Majoris obtained after the second phase self
calibration iteration. We measure 
$\rho=143.89$~mas and $\theta=306.9^{\circ}$, in good agreement with the 
ephemerides of this source~\cite{Mason06}.} 
   \end{figure}
   \begin{figure}
   \begin{center}
   \begin{tabular}{c}
   \includegraphics[height=9.5cm]{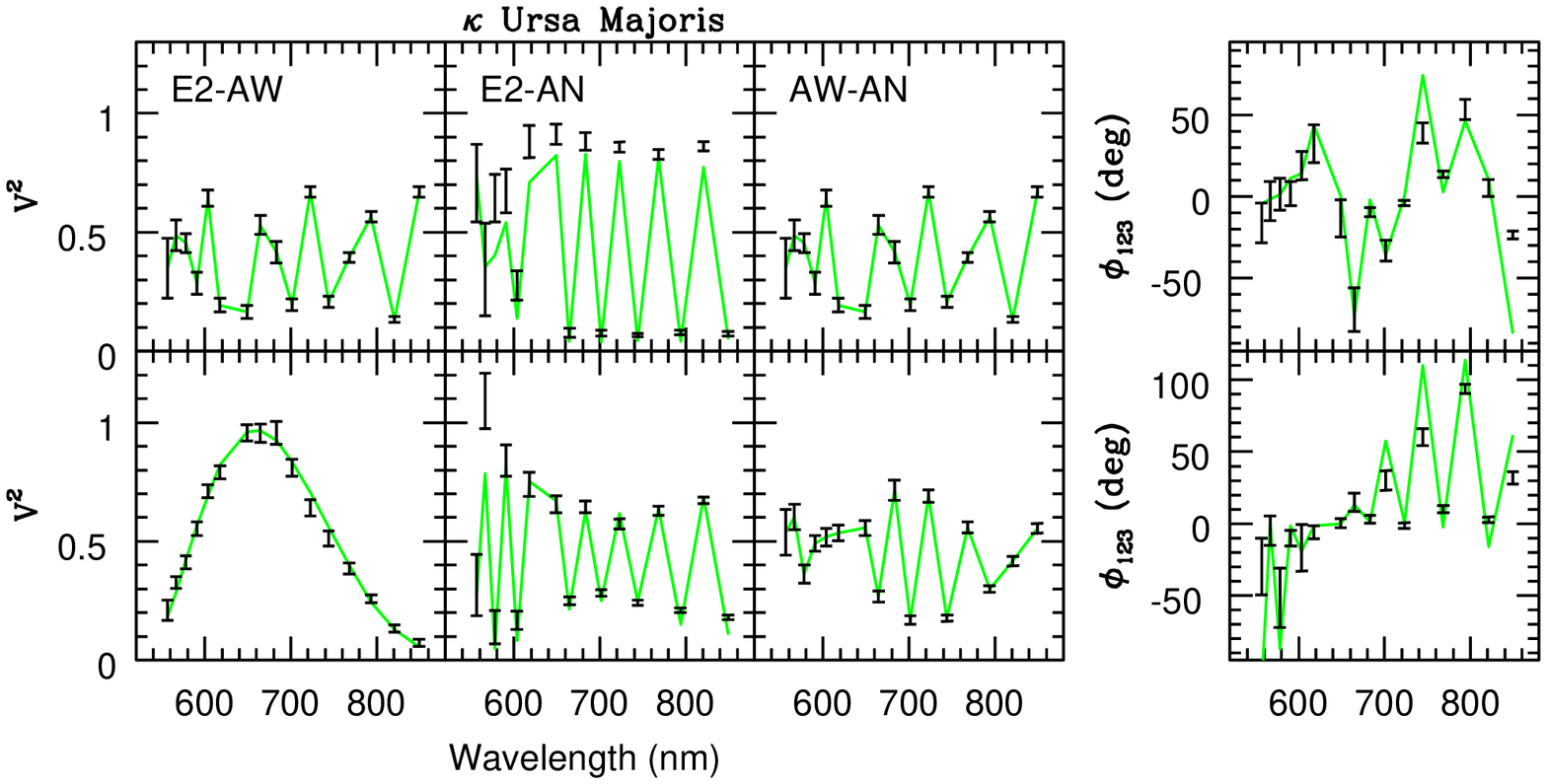}
   \end{tabular}
   \end{center}
   \caption[example]
   {Measured V$^2$'s and triple phases of two scans of
$\kappa$ Ursa Majoris are shown as error bars. The solid line shows the
model calculated using the parameters measured in the image presented
in Figure~3.}
   \end{figure}

We show in Figure~2 the results of this technique applied to the
observations of the binary system $\zeta$ Draconis. The Figure shows
the image obtained after one iteration cycle of phase self calibration,
created using all the channels. The measured separation and position
angle between the 2 sources agress with the values obtained from the $V^2$
fits. Given the width of our channels and the separation between the 2 stars,
bandwidth smearing causes a noticeable reduction in the flux of the
fainter source, which should be taken into account in the measurements.
The dynamic range of this image is $\sim$100. Although
we could do additional phase self calibration iterations, we find that
in this case the improvement to the image quality is not significant.

In Figure~3 we show the image of the binary system $\kappa$ Ursa Majoris,
obtained after two phase self calibration iteration cycles. We measure
the separation and position angle between the two sources, which agree
very well with the published ephemerides of this source.
As an alternative confirmation of these results we use the parameters
measured from this image to create a model and calculate the expected
$V^2$'s and triple phases of this system. The results from this model are
compared to the observed $V^2$'s and triple phases in
Figure~4, where we can see that there is a good agreement between the model
and the measurements.

In Figure~5 we show the application of this technique to a more
complicated source, the triple system $\eta$ Virginis. The image shows
the inner binary, separated by 5.53~mas along p.a.$=306.1^{\circ}$,
while the outer component is separated by 109.03~mas along
p.a.$=199.0^{\circ}$, relative to the photocenter of the inner
pair. These measurements agree with the ephemerides published by
Ref.~\citenum{Hummel03a}.

Finally, in Figure~6 we show another 2 binary systems, $\theta^2$ Tauri and
$\phi$ Herculis. Again, the separations and position angles
of the components agree with their published ephemerides~\cite{Armstrong06,Zavala07}.

   \begin{figure}
   \begin{center}
   \begin{tabular}{c}
   \includegraphics[height=12cm]{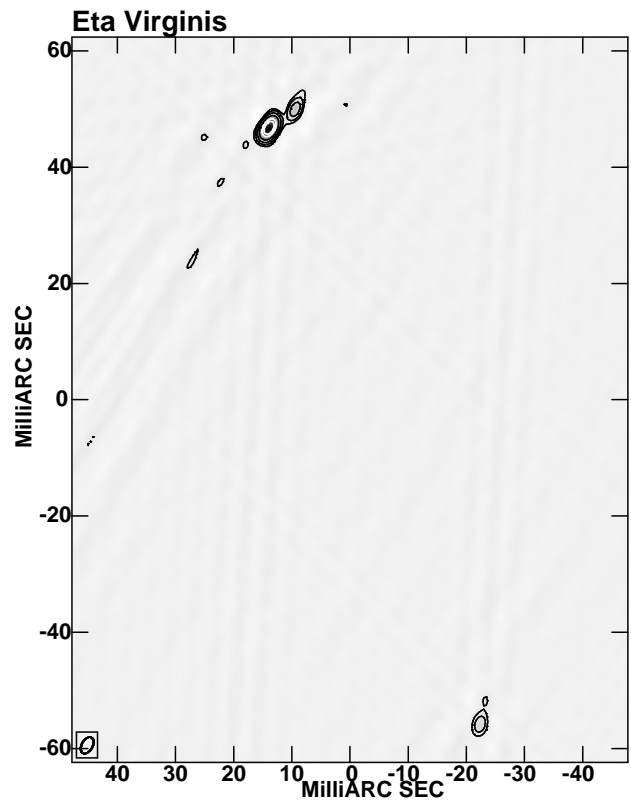}
   \end{tabular}
   \end{center}
   \caption[example]
   {Image of $\eta$ Virginis obtained after the second phase self calibration iteration.}
   \end{figure}
   \begin{figure}
   \begin{center}
   \begin{tabular}{c}
   \includegraphics[height=6cm]{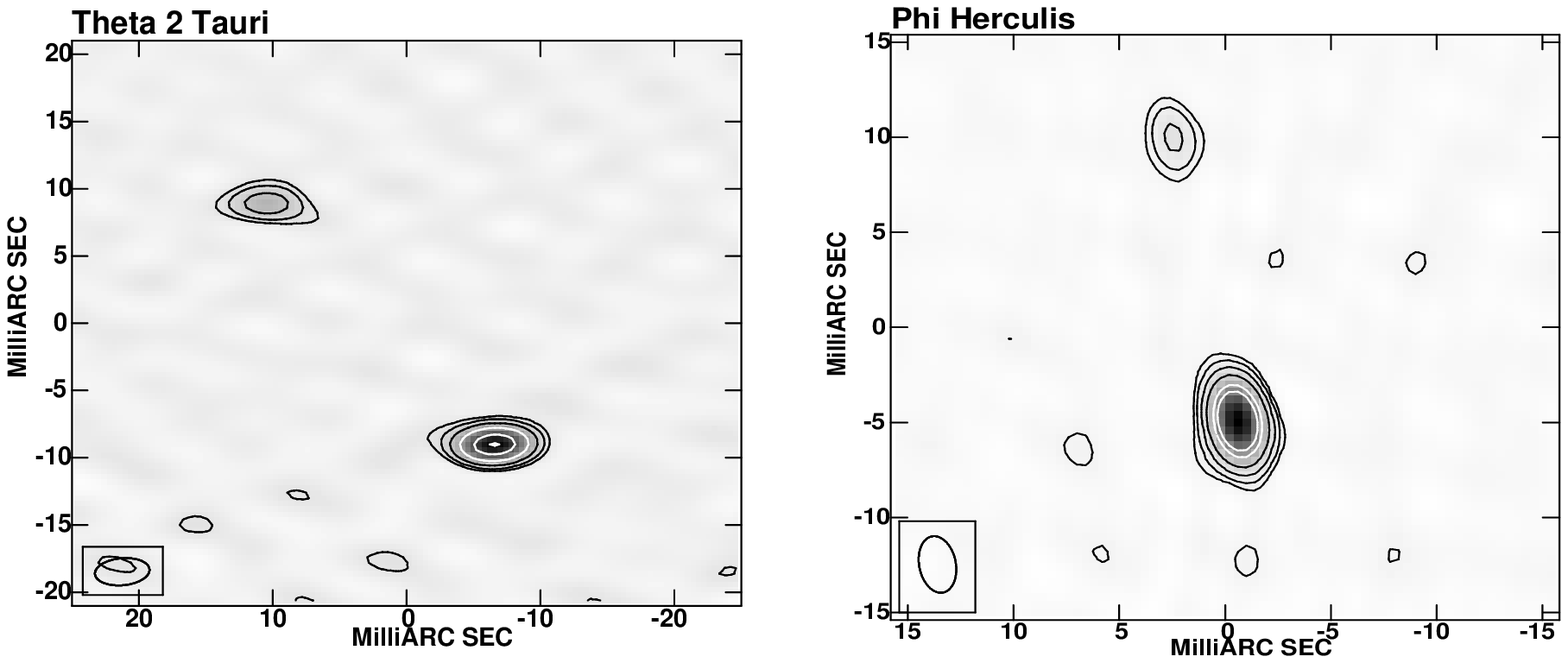}
   \end{tabular}
   \end{center}
   \caption[example]
   {Images of $\theta^2$ Tauri (left) and $\phi$ Herculis (right), obtained
after one phase self-calibration iteration. }
   \end{figure}

\section{SUMMARY} 

In this paper we discussed the development of two techniques that
used coherently averaged NPOI data to recover complex visibilities.
This allowed us to image the sources using standard radio interferometry
imaging methods. Using the first technique, differential phases, we
were able to correct the instrumental and atmospheric contribution
to the intrinsic H$\alpha$ phases of $\beta$ Lyrae, being able to image
the line emission in this source and detect its movement relative to
the continuum photocenter. Although this is a powerful technique,
it requires a priori knowledge about the structure of the source at
some wavelength, and has a limited number of applications. The second
technique, self calibration, was used to correct the atmospheric effects
to the observed phases, allowing us to image binary and triple systems.
This technique is extensively used in radio interferometry, requires
closure phases, but is applicable to a wider range of sources. Future
plans include the application of this technique to other sources, like
rapidly rotating stars, stars with spots and disks, among other sources.

We would like to point out that the techniques presented here
make use of all the information collected by the interferometer and do
not combine multiple measurements (e.g. closure phases), like in the
case of imaging methods that use only $V^2$'s and closure phases.
Consequently we should be able to obtain higher precision measurements
and/or higher dynamic range images of astrophysical sources.

\acknowledgments     
 
The work done with the NPOI was performed through a collaboration between
the Naval Research Laboratory and the US Naval Observatory, in association with
Lowell Observatory, and was funded by the Office of Naval Research and the
Oceanographer of the Navy. Basic research in optical interferometry at the NRL
is supported by 6.1 Base funding. We thank the NPOI staff for the careful
observations that contributed to this work. This research has made use
of the SIMBAD literature database, operated at CDS, Strasbourg, France.
We would like to thank Christian Hummel for the availability of the 
data reduction package OYSTER.


\bibliography{report}   
\bibliographystyle{spiebib}   

\end{document}